\documentclass[aps,prl, twocolumn]{revtex4-1}
\usepackage{amsmath, amssymb, graphicx, color}

\def\aap{Astronomy \& Astrophysics}
\def\aj{Astronomical Journal}

\def\mnras{MNRAS}
\def\jcap{JCAP}
\def\prd{Physical Review D}
\def\prl{Physical Review Letters}
\def\apj{Astrophysical Journal}

\def\epss{\varepsilon_s}
\def\dom{\delta_\Omega}
\def\aeq{a_{\rm eq}}
\def\omm{\Omega_\mathrm{m}}
\def\mpl{M_\mathrm{Pl}}
\def\wde{w_\mathrm{DE}}
\def\tildeomm{\widetilde{\Omega}_\mathrm{m}}

\begin{document}

\title[ALAverse]{Falsifiable anthropic model from the string landscape}
\author{Zhiqi Huang}
\email{huangzhq25@mail.sysu.edu.cn}
\affiliation{School of Physics and Astronomy, Sun Yat-sen University, 2 Daxue Road, Zhuhai, 519082, China}
\affiliation{CSST Science Center for the Guangdong-Hong Kong-Macau Greater Bay Area, Zhuhai, 519082, China}

\begin{abstract}
  We combine the axiverse with a negative cosmological constant $\Lambda$, both originating from the string landscape, and apply an observation-time-weighted anthropic argument. Adopting a uniform prior on $\Lambda$ and a typical string-motivated axion decay constant comparable to the reduced Planck scale, this framework predicts a $\sim 40\%$ probability of observing $0.1 < \Omega_m < 0.9$, thereby naturally resolving the long-standing fine-tuning and coincidence problems of dark energy. Unlike many scalar-field dark energy models, the anthropic Lambda-axion universe (ALAverse) statistically disfavors slow-roll dynamics, as slow-roll requires fine-tuning of the initial field displacement. Moreover, the negative cosmological constant renders fast-roll scenarios anthropically unfavorable, since they typically yield only a very brief observational window with positive dark energy density. Having ruled out both extremes, the ALAverse characteristically predicts a moderate-roll dynamics. We derive a two-parameter parametrization in terms of $\dom$ and $|\epss|$ that covers ALAverse solutions as well as a broad class of canonical and phantom field models. Current observational data yield $\dom = -0.0498 \pm 0.0186$, corresponding to a $2.7\sigma$ rejection of $\Lambda$CDM ($\dom = 0$) and  single-field thawing phantom models  ($\dom > 0$). The data also show a mild preference for the ALAverse over slow-roll quintessence, a trend that can be conclusively tested with future high-precision measurements of the Hubble diagram.
\end{abstract}


\maketitle

\section{Introduction}\label{sec:intro}

Since the discovery of late-time cosmic acceleration in 1998~\cite{Riess1998, Perlmutter1999}, the concordance cosmology has been remarkably successful in explaining a plethora of observational data~\cite{Planck2018Params}. Nevertheless, three fundamental ingredients of the model remain poorly understood: early-universe inflation, dark matter, and dark energy. In the minimal six-parameter Lambda cold dark matter ($\Lambda$CDM) framework, dark energy is simply parameterized as a cosmological constant $\Lambda$ (i.e., the vacuum energy), while inflation and dark matter are treated only phenomenologically.

Despite its observational triumphs, this interpretation of dark energy faces severe theoretical challenges. In the presence of unbroken supersymmetry, the vacuum energy is exactly zero. Yet to accommodate a universe like ours, supersymmetry must be broken at some energy scale $M_s$ well beyond the reach of current colliders ($\sim \mathrm{TeV}$). A naive dimensional analysis then suggests a vacuum energy density at least of order $M_s^4 \gtrsim \mathrm{TeV}^4$, which exceeds the observed dark energy density by over $60$ orders of magnitude. This is the well-known cosmological constant problem~\cite{Weinberg1989}, or fine-tuning problem of dark energy.

String theory predicts a vast number of vacua with negative vacuum energy density. To obtain a universe with a positive cosmological constant, one must carefully construct metastable vacua with sufficiently long lifetimes~\cite{KKLT}. Within the string landscape, the anthropic principle~\cite{Carter1974} then provides a natural selection argument: a cosmological constant whose magnitude is too large---whether positive or negative---would inhibit galaxy formation and hence preclude the emergence of intelligent observers~\cite{Weinberg1987}. However, this bare anthropic argument does not explain why the cosmological constant is positive, nor why its magnitude happens to be so close to the matter density at the epoch when intelligent observers emerge. This is known as the coincidence problem of dark energy~\cite{Zlatev1999}.

String theory also predicts the existence of a large number of ultralight axions, with masses potentially extending down to the Hubble scale $\sim 10^{-33}\,\mathrm{eV}$. The heavier ones, with masses $m \gtrsim 10^{-28}\,\mathrm{eV}$, are viable dark matter candidates, while the lighter ones with $m \sim 10^{-33}\,\mathrm{eV}$ offer a natural candidate for dark energy. This scenario is known as the string axiverse~\cite{Arvanitaki2010}. An even more speculative conjecture is that early-universe inflation may also be driven by string axions~\cite{Freese1990, Odintsov2019}.

Given that stable vacua with a negative cosmological constant are far more generic in the string landscape than metastable vacua with a positive cosmological constant and sufficiently long lifetimes~\cite{Danielsson2018}, one may conjecture that the bare vacuum energy density in our universe is in fact negative. Within the string axiverse framework, the observed positive dark energy density can then be interpreted as the sum of this negative bare vacuum energy and the energy density of an ultralight axion field. One may further apply an anthropic selection argument to this $\Lambda$-axion scenario. A recent work by Murai et al.~\cite{Murai2025} suggests that, when an anthropically motivated upper bound is imposed on the total dark energy density, and under an additional assumption regarding the onset of late-time cosmic acceleration, the axion mass is typically of order $m = \mathcal{O}(10)\,H_0$ for a decay constant near the Planck scale, where $H_0$ is the Hubble constant. This prediction, however, is in tension with the constraint $m \sim H_0$ derived from recent observational data~\cite{Luu2025a}. Moreover, the additional assumption on the onset of cosmic acceleration is not a genuinely anthropic argument; rather, it amounts to accepting the coincidence problem as a prior.

Despite its popularity, the anthropic principle is rarely regarded as a falsifiable theory. The core difficulty lies in the fact that different schemes for assigning probabilities to candidate universes can yield drastically different anthropic predictions for the cosmological constant~\cite{Starkman2006, Sudoh2017}. This ambiguity ultimately stems from our inability to evaluate the number density of intelligent life---a concept that is not even well-defined in our own universe---let alone in universes that are radically different from ours. Thus, it is unlikely that any anthropic argument can unambiguously resolve the coincidence problem. Nevertheless, we may at least adopt a minimal consistency criterion: an anthropic argument should be deemed unhelpful if it either presupposes a resolution to the coincidence problem~\cite{Murai2025}, or worse, yields predictions that are in direct tension with it~\cite{Starkman2006, Sorini2024}.

In the present work, we apply a simple observation-time-weighted anthropic argument to the $\Lambda$-axion scenario. To our knowledge, this weighting scheme has not been explored in previous studies. In conventional anthropic settings with a positive cosmological constant, the observation time available to intelligent life is enormously prior-dependent, as metal-rich galaxies may survive for up to trillions of years. By contrast, in a $\Lambda$-axion universe, whose ultimate fate is a big crunch, anthropic selection typically predicts a cosmic lifetime of a few tens of billions of years~\cite{Luu2025b}. This makes the observation-time weighting both well-defined and computationally tractable. Our working principle is simple: among universes that satisfy the basic conditions for habitability, we assign greater weight to those that offer a longer cumulative duration for observation. We do not claim that our weighting scheme is superior to all possible alternatives~\cite{Efstathiou1995, Garriga1999, Peacock2007, Bousso2010, Barnes2018, Salcido2018, Oh2022}; rather, we emphasize that it yields explicit, falsifiable predictions that can be tested with current and upcoming observational data.

Throughout this paper, we work in natural units with $c = \hbar = 1$ and adopt a spatially flat Friedmann--Lema\^{\i}tre--Robertson--Walker metric, $\mathrm{d}s^2 = \mathrm{d}t^2 - a^2(t)\,\mathrm{d}\mathbf{x}^2$, where $t$ is the cosmic time and $a(t)$ is the scale factor. The Hubble parameter is defined as $H \equiv \dot{a}/a$, where an overdot denotes differentiation with respect to $t$. A subscript ``0'' denotes quantity today, and unless otherwise specified, the scale factor is normalized to unity today ($a_0\equiv 1$). The Hubble constant $H_0$, i.e., the Hubble parameter today, is sometimes expressed with the reduced Hubble constant $h$ via $H_0 = 100h\,\mathrm{km/s/Mpc}$.

\section{Theoretical framework and anthropic predictions}\label{sec:theory}

We consider a classical scalar axion $\phi$ with a potential
\begin{equation}
  V(\phi) = m^2 f^2 \left(1 + \cos \frac{\phi}{f}\right),
\end{equation}
where $m$ is the axion mass and $f$ is the decay constant. To make this model a viable candidate for dark energy, the decay constant must be close to the Planck scale. For simplicity, we fix $f = \mpl \equiv 1/\sqrt{8\pi G}$, where $G$ is Newton's gravitational constant and $\mpl$ is the reduced Planck mass. The negative cosmological constant $\Lambda$ can be effectively absorbed as an additional constant term $\rho_\Lambda \equiv \Lambda / (8\pi G)$ in the effective axion potential,
\begin{equation}
  V_{\rm eff}(\phi) = m^2 f^2 \left(1 + \cos \frac{\phi}{f}\right) + \rho_\Lambda. \label{eq:Veff}
\end{equation}
The $\Lambda$-axion model is therefore specified by three parameters: the axion mass $m$, the negative vacuum energy $\rho_\Lambda$ (or equivalently $\Lambda$), and the initial field displacement $\phi_i$.

We fix the early-universe conditions—namely, the baryon-to-photon ratio, the dark-matter-to-baryon ratio, and the primordial power spectra of curvature fluctuations—while keeping all other fundamental parameters fixed except for the three dark-energy parameters above. This is not the typical situation in the string landscape, where essentially all fundamental parameters may vary. Nevertheless, our philosophy here is to focus specifically on the anthropic selection of $\Lambda$ while keeping the statistical framework as simple and transparent as possible.

In the string axiverse picture, axion masses are exponentially sensitive to compactification parameters and are expected to be distributed across many orders of magnitude rather than clustered around a single scale~\cite{Arvanitaki2010}. We accordingly adopt a uniform prior in $\ln m$. We will show, however, that our results remain qualitatively unchanged if a uniform prior in $m$ is used instead. For $\rho_\Lambda$ and $\phi_i$, we assume flat (uniform) priors.

We split the analysis into two steps. In the first step, we randomly draw an observation time within a specified anthropic window (to be defined below) and compute the corresponding matter density parameter $\omm$. The coincidence problem can then be quantified, for instance, by the probability of finding $0.1 < \omm < 0.9$. In the second step, we focus on the subset of solutions with $\omm \simeq 0.3$---i.e., those compatible with our actual universe---and extract the distinctive imprints left on the time evolution of the dark energy equation of state (EOS), $\wde(z) \equiv p_{\rm DE}(z)/\rho_{\rm DE}(z)$, where $p_{\rm DE}$ and $\rho_{\rm DE}$ are the pressure and energy density of the dark energy component, respectively. 

\subsection{Anthropic selection of $\omm$}

In most anthropic arguments, different universes are weighted according to the number of intelligent observers they contain. However, the density of intelligent observers per unit spacetime volume is far from straightforward to quantify. The situation is further complicated by the fact that observers possess memory and can communicate with one another. Even within our own case, it is unclear whether the "surpriseness" about the value of $\Lambda$ should be multiplied by the number of cosmologists, the number of human beings, or perhaps the total number of observers over the entire history of humanity. It therefore appears hopeless to construct a well-defined, observer-based weighting scheme. We thus adopt a different philosophy: observations made at the same cosmic time should not be double counted. This simplifies the problem to a binary choice—namely, whether intelligent life can exist under a given cosmological environment.

It is commonly assumed that intelligent observers reside in large, metal-rich galaxies similar to our Milky Way, whose associated dark matter halos collapsed from comoving volumes of order $\sim \mathrm{Mpc}^3$. We therefore define the onset of the anthropic window as the moment when $\sigma_{\mathrm{1\,Mpc}}$---the amplitude of linear matter density fluctuations $\delta_m$ smoothed over a comoving sphere of radius $1\,\mathrm{Mpc}$---exceeds the spherical-collapse threshold of $1.686$~\cite{Gunn1972, Press1974}. This condition typically signals that objects collapsing from $\sim\!\mathrm{Mpc}$-scale volumes have virialized. In the $\Lambda$-axion model, the negative cosmological constant ensures that every universe ultimately ends in a big crunch. We accordingly define the termination of the anthropic window as the time when the effective potential given by Eq.~\eqref{eq:Veff} becomes negative. 

The above estimate of the anthropic window is admittedly crude. In principle, refined calculations incorporating detailed star-formation histories and the survival time of intelligent life prior to the crunch could improve the accuracy. Such efforts, however, would require complicated astrophysical modeling and introduce additional assumptions, with uncertainties that may reach several billion years~\cite{Peacock2007, Sorini2024}. Nevertheless, the typical anthropic window in the $\Lambda$-axion model spans a few tens of billions of years, which is sufficiently long that astrophysical uncertainties of a few billion years can be regarded as small relative corrections. Indeed, we do not need an exact probability for, say, $0.1<\omm<0.9$; an order-of-magnitude estimate suffices to address the coincidence problem.

Fine-tuned initial conditions could, in principle, make the anthropic window defined above far longer than typical stellar lifetimes. To ensure that the random sampling procedure remains computationally practical, we therefore impose a conservative cutoff of a trillion years on the anthropic window. We stress, however, that this cutoff has little impact on our statistical results: any case that would yield a significantly broader window would require extreme fine-tuning of the initial conditions, which is already statistically suppressed by the prior.

For each randomly sampled set of parameters, we choose the initial conditions in the matter dominated regime and evolve the background equations
\begin{eqnarray}
  \rho_{\rm m} &=& \rho_{\rm m,ini}\left(\frac{a}{a_{\rm ini}}\right)^{-3}, \label{eq:rho_m} \\
  \frac{\ddot a}{a} + \frac{1}{6\mpl^2}\left[2\dot\phi ^2 - 2V_{\rm eff}(\phi)+\rho_{\rm m}\right] &=& 0, \label{eq:Rayleigh} \\
  \ddot\phi + 3H\dot\phi + \frac{\mathrm{d}V_{\rm eff}}{\mathrm{d}\phi} &=& 0, \label{eq:KG}
\end{eqnarray}
supplemented by the linear growth equation for matter density perturbations,
\begin{equation}
  \ddot \delta_{\rm m} + 2H\dot\delta_{\rm m} - \frac{\rho_{\rm m}}{2\mpl^2}\delta_{\rm m} = 0. \label{eq:growth}
\end{equation}

Here $\rho_{\rm m}$ is the background matter density and $\delta_{\rm m}$ is its relative perturbation.
The Friedmann constraint,
\begin{equation}
  3H^2\mpl^2 = V_{\rm eff}(\phi) + \frac{1}{2}{\dot\phi}^2 + \rho_{\rm m}, \label{eq:Friedmann}
\end{equation}
is used to set the initial conditions and to monitor numerical accuracy. The remaining initial conditions ($\rho_{\rm m, ini}$ and the primordial power spectrum of $\delta_{\rm m}$) are set according to the Planck best-fit cosmology~\cite{Planck2018Params}. Note that we have neglected field perturbations $\delta\phi$, which are safely negligible on the deeply subhorizon $1\,\mathrm{Mpc}$ scales relevant to our analysis.

We run Markov chain Monte Carlo (MCMC) to sample cosmologies within the following ranges: $\log_{10}(m / H_{\mathrm{0,ref}}) \in [-30, 30]$, where the reference Hubble scale is $H_{\mathrm{0,ref}} = 67.3\,\mathrm{km\,s^{-1}\,Mpc^{-1}}$; $\rho_\Lambda/\mathrm{eV}^4 \in [-10^{30}, 0]$; and $\phi_i/f \in [0, \pi]$. Additionally, in order to match the observed primordial power spectrum and dark matter density, we impose that the initial dark energy density is subdominant to the dark matter density; parameter combinations that violate this condition are rejected before entering the likelihood evaluation. The sampling procedure is designed to weight each cosmology proportionally to the duration of its anthropic window. Specifically, we employ a Metropolis--Hastings algorithm in which the likelihood function for a given set of cosmological parameters is defined as the length of the anthropic window, i.e., $\mathcal{L} \propto \Delta t_{\rm obs}$, where $\Delta t_{\rm obs}$ is the total time during which the universe admits intelligent observers. This choice ensures that cosmologies with longer anthropic windows are preferentially sampled in the Markov chain, with an acceptance probability that automatically accounts for the relative window durations. For each sampled cosmology in the chain, we then draw a single observation time uniformly within its anthropic window. This two-step procedure—first weighting the cosmology by its window length, then conditioning on a random observation time within that window—is statistically equivalent to drawing observation times from the full ensemble of cosmologies with a probability proportional to the duration of each window. In other words, the weighting by observation time is already encoded in the MCMC acceptance ratio; the subsequent uniform draw merely selects a representative epoch for computing quantities such as $\Omega_m$ without introducing additional bias. This approach avoids the need to draw multiple observation times per cosmology, which would be computationally prohibitive, while correctly implementing the observation-time-weighted anthropic principle.

\begin{figure}
  \includegraphics[width=0.48\textwidth]{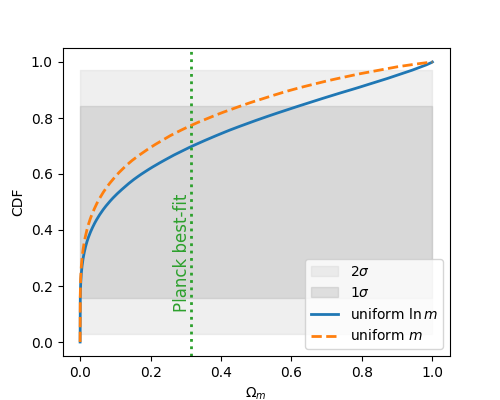}
  \includegraphics[width=0.48\textwidth]{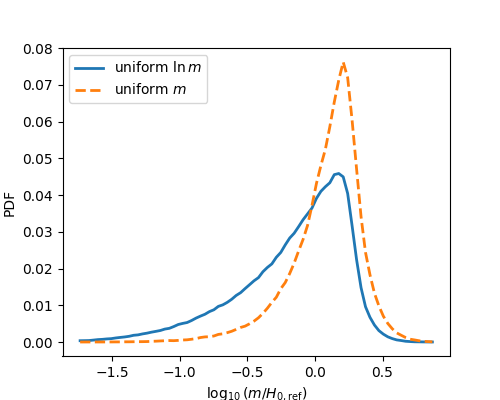}
  \caption{Anthropically selected CDF of $\omm$ (upper) and distribution of $\log_{10}(m / H_{\mathrm{0,ref}})$ (lower), where $H_{\mathrm{0,ref}} = 67.3\,\mathrm{km\,s^{-1}\,Mpc^{-1}}$. \label{fig:cdf}}
\end{figure}

The upper panel of Figure~\ref{fig:cdf} shows the cumulative distribution function (CDF) of $\omm$. It demonstrates that the value of $\omm$ observed in our own universe lies well within the typical region of the predicted distribution. For comparison, the probability of finding $0.1 < \omm < 0.9$ is $43\%$ under a uniform prior in $\ln m$, and $39\%$ under a uniform prior in $m$. The coincidence between dark energy and matter densities is thus a natural outcome in the ALAverse, rather than a cause for surprise.

To understand how the anthropic argument resolves the coincidence problem, it is instructive to examine why the axion mass $m$ is effectively bounded by anthropic selection---neither too large nor too small. The lower panel of Figure~\ref{fig:cdf} shows that the anthropically allowed region of $m$ is confined to within roughly one decade around the reference Hubble scale. The upper bound is straightforward: a larger $m$ tends to suppress the growth of matter density perturbations and to drive the axion into the fast-roll regime; both effects shorten the anthropic window. The lower bound is more subtle. It arises from the requirement that the initial dark energy density---i.e., the sum of the axion potential and the randomly drawn (negative) cosmological constant term $\rho_\Lambda$---must be positive. Since $\rho_\Lambda$ is independently sampled from a flat prior, it cannot be adjusted to accommodate arbitrarily small $m$; hence, if $m$ is too small, the negative $\rho_\Lambda$ would be more likely to dominate and make the total dark energy density negative from the outset. We have neglected the rare possibility that a negative but very small total dark energy density could still admit an anthropic window of a few billion years~\cite{Peacock2007}. This simplification is justified because such cases are statistically rare, whereas the typical anthropic window is much longer, of order several tens of billions of years~\cite{Luu2025b}. 

\subsection{Dark Energy EOS \label{sec:eos}}

We now focus on solutions that resemble our own universe—namely, those with $\omm \simeq 0.3$ and dark energy driving the late-time cosmic acceleration. To isolate such solutions, we impose a Gaussian prior $\omm = 0.315 \pm 0.01$ and require $\wde < -1/3$ in the sampling procedure. To extract statistical information and facilitate comparisons between the ALAverse and a broader class of scalar-field dark energy models, we seek a phenomenological parametrization of the dark energy EOS that encompasses a wide range of scenarios, including the ALAverse.

In the slow-roll limit ($1+\wde \ll 1$), the evolution of $1+\wde$ for a canonical scalar field in the presence of matter can be well approximated by~\cite{HBK2011, MH2018, Huang2021}
\begin{equation}
  1+\wde(a) \approx \epss \, g\!\left(\frac{a^3}{\aeq^3}\right), \label{eq:1param}
\end{equation}
where $\aeq$ is the scale factor at matter--dark energy equality, and the slow-roll evolution function $g$ is given by
\begin{equation}
  g(x) \equiv \frac{2}{3}\left[\sqrt{1+\frac{1}{x}} - \frac{\ln\left(\sqrt{x} + \sqrt{1+x}\right)}{x}\right]^2. \label{eq:fdef}
\end{equation}
The parameter $\epss$, defined as
\begin{equation}
  \epss\equiv \frac{1+\wde(\aeq)}{g(1)} \approx 5.2832\left[1+\wde(\aeq)\right], \label{eq:epssdef}
\end{equation}
can be viewed as a normalization factor quantifying the deviation from $\Lambda$CDM, with $\epss=0$ recovering the cosmological constant. In the slow-roll limit, $\epss$ is related to the scalar potential via~\cite{HBK2011}
\begin{equation}
  \left.\epss\right\vert_{\rm slow-roll} \approx \frac{\mpl^2}{2}\left(\frac{\mathrm{d}\ln V_{\rm eff}}{\mathrm{d}\phi}\right)^2\bigg|_{a=\aeq}, \label{eq:epsv}
\end{equation}
which is manifestly nonnegative for a canonical scalar field. More generally, Eq.~\eqref{eq:1param} can be extended to phantom fields by allowing $\epss$ to take negative values~\cite{HBK2011}.

For a given $\omm$ and any parametrization of $\wde$, the definition of $\aeq$ implies $\rho_{\rm DE}=\rho_m$ at $a=\aeq$. By integrating the energy conservation equations from $a=\aeq$ to $a=1$, we obtain a consistency relation
\begin{equation}
  (1-\Omega_m)\exp\!\left[\int_{\aeq}^1 \frac{3(1+\wde(a))}{a}\,\mathrm{d}a\right] =  \Omega_m \aeq^{-3}. \label{eq:consis}
\end{equation}
For parametrization~\eqref{eq:1param}, the above consistency relation leads to
\begin{equation}
  \left.\epss \, F(\aeq^{-3})\right\vert_{\rm slow-roll} = \ln\frac{\omm}{(1-\omm)\,\aeq^3}, \label{eq:1param_consis}
\end{equation}
where the function $F$ is defined as
\begin{equation}
  F(\mu) \equiv \int_1^\mu \frac{g(x)}{x}\,\mathrm{d}x. \label{eq:Fdef}
\end{equation}
Since $\omm$ is usually treated as a separately measured cosmological parameter rather than a free parameter of the dark energy model, Eq.~\eqref{eq:1param} effectively constitutes a one-parameter parametrization of dark energy. In the original work~\cite{HBK2011}, $\epss$ was taken as the fundamental parameter, with $\aeq$ derived from it via an approximate explicit expression—sufficient for the data accuracy at that time. For today's high-precision observations, however, we adopt the exact consistency relation~\eqref{eq:1param_consis} to ensure that our results are not biased by crude approximations.

Parametrization~\eqref{eq:1param} requires extension beyond the slow-roll regime when applied to the moderate-roll solutions that are more common in the ALAverse. The original work~\cite{HBK2011} extended the one-parameter parametrization~\eqref{eq:1param} by introducing two additional parameters: $\epsilon_{\phi\infty}$, which describes the field velocity at high redshift, and $\zeta_s$, which describes the low-redshift variation of the field velocity. This three-parameter ansatz covers a much broader class of scalar-field models, including ``thawing models'' (where the field is frozen by Hubble friction at high redshift) and ``freezing models'' (where the field follows a fast-roll scaling solution at high redshift). The scaling solutions require a very steep potential to overcome Hubble friction, a feature clearly absent in the effective potential $V_{\rm eff}$ of the ALAverse. Slow-roll ALAverse solutions therefore fall into the category of thawing models, for which $\epsilon_{\phi\infty}=0$ and $1+\wde$ is parametrized by $\epss$ and $\zeta_s$. In thawing scenario, the original $\zeta_s$ correction implicitly relies on the assumption that the scalar potential does not approach zero. In the ALAverse, however, we wish to include the possibility that $V_{\rm eff}$ approaches $0^+$ due to the negative cosmological constant. In that case, the logarithmic derivative $(\mathrm{d}\ln V_{\rm eff}/\mathrm{d}\phi)^2$ can rise sharply, giving rise to an accelerated-thawing behavior that cannot be adequately captured by the $\epss$-$\zeta_s$ approximation. This limitation motivates us to introduce a new parametrization that accommodates such rapid-rising features—a direction further motivated by recent baryon acoustic oscillation (BAO) measurements from the Dark Energy Spectroscopic Instrument (DESI), which indicate a possible rapid rise in $1+\wde$ at low redshifts~\cite{DESIDR2}.

We therefore propose the following extension to Eq.~\eqref{eq:1param}:
\begin{equation}
  1+\wde \approx \epss \, g\!\left(\frac{a^3}{\aeq^3}\right)\exp\!{\left[\lambda F\!\left(\frac{a^3}{\aeq^3}\right)\right]}, \label{eq:newparam}
\end{equation}
where
\begin{equation}
  \lambda \equiv \frac{1}{F(\aeq^{-3})}\, B\!\left(\frac{1}{\epss F(\aeq^{-3})}\ln\frac{\omm}{(1-\omm)\aeq^3}\right), 
\end{equation}
with $B$ the inverse function of $(e^x-1)/x$, and $g$ and $F$ defined in Eqs.~\eqref{eq:fdef} and~\eqref{eq:Fdef}, respectively.

In the slow-roll limit, the new parametrization~\eqref{eq:newparam} can be derived analytically by iteratively solving the slow-roll equations of motion, and may therefore be viewed as a second-order extension of Eq.~\eqref{eq:1param}. One can further show that the factor in the exponential is related to the scalar potential via
\begin{equation}
\lambda \vert_{\rm slow-roll}
  \approx -\mpl^2\frac{\mathrm{d}^2\ln V_{\rm eff}}{\mathrm{d}\phi^2}, \label{eq:lamslowroll}
\end{equation}
which explicitly involves a second-order expansion of the logarithmic potential. 

From the perspective of energy conservation, parametrization~\eqref{eq:newparam} goes beyond a second-order approximation. It satisfies the consistency equation~\eqref{eq:consis} for arbitrary choices of $\aeq$ and $\epss$, as long as the argument of $B$ is positive. That means Eq.~\eqref{eq:newparam} is a two-parameter (namely $\epss$ and $\aeq$) dark energy parametrization with {\it exact} energy conservation between $a=\aeq$ and $a=1$. This feature makes parametrization~\eqref{eq:newparam} a highly accurate approximation for slow-roll and moderate-roll thawing models.

Since both $\aeq$ (the scale factor at $\rho_{\rm DE}=\rho_m$) and $\epss$ (defined by Eq.~\eqref{eq:epssdef}) can be directly read out from numerical solutions, we may directly compare the exact solution of $1+\wde$ with the approximation given by Eq.~\eqref{eq:newparam}. 

\begin{figure*}
  \includegraphics[width=\textwidth]{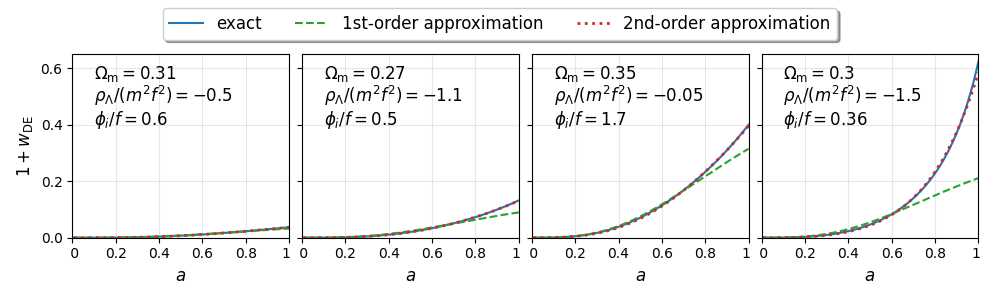}
  \caption{Dark energy EOS from ALAverse numerical solutions and comparison with the first-order approximation~\eqref{eq:1param} and the second-order approximation~\eqref{eq:newparam}.}\label{fig:wfit}
\end{figure*}

Figure~\ref{fig:wfit} compares a few numerical solutions with the first-order approximation~\eqref{eq:1param} and parametrization~\eqref{eq:newparam}. In all cases, Eq.~\eqref{eq:newparam} shows excellent agreement with sub-percent accuracy. Numerical experiments on many other thawing models indicate that parametrization~\eqref{eq:newparam} is universally superior to the original $\epss$-$\zeta_s$ approximation, which typically produces percent-level errors when $1+\wde\gtrsim 0.2$~\cite{HBK2011}.

The positivity condition on the argument of $B$ in parametrization~\eqref{eq:newparam} translates into a nontrivial boundary in the $\epss$-$\aeq$ space. To make the parameter space more tractable, we define
\begin{equation}
  \tildeomm \equiv \frac{1}{1+\aeq^{-3}},
\end{equation}
which in $\Lambda$CDM coincides with $\omm$. The relative difference between $\tildeomm$ and $\omm$ defines a new variable that combines both $\omm$ and $\aeq$:
\begin{equation}
  \dom \equiv \frac{\tildeomm}{\omm}-1.
\end{equation}
This quantity therefore encodes the deviation from $\Lambda$CDM. For a canonical scalar field, we have $\dom \le 0$ and $\epss \ge 0$; for the phantom case, the opposite holds ($\epss \le 0$ and $\dom \ge 0$). In other words, upon replacing $\aeq$ by $\dom$, the positivity of the argument of $B$ simply reduces to $\epss\,\dom \le 0$; { while $\dom=0$ automatically implies $\epss=0$ and thus corresponds to the $\Lambda$CDM model.} It is therefore more convenient to use $|\epss|$ and $\dom$ as the fundamental parameters, with $\aeq$ treated as a derived quantity.

The sampled ALAverse solutions are then mapped into the $\dom$-$|\epss|$ space. The resulting distribution is to be compared with the observational constraints presented in the following section.

\subsection{Robustness of the anthropic window endpoint}

In the main analysis, we defined the end of the anthropic window as the time when the effective potential vanishes, i.e., $V_{\rm eff}=0$. However, as pointed out in Ref.~\cite{Peacock2007}, a negative cosmological constant might still admit a brief anthropic window of a few billion years before the big crunch. Furthermore, in our model the dark energy density includes a positive kinetic component, so $V_{\rm eff} < 0$ does not necessarily imply a negative total dark energy density. Finally, a contracting phase ($H<0$) does not necessarily imply a negative dark energy density either, since the axion field can roll up the potential again during contraction. These subtleties motivate a robustness check on the choice of the anthropic window endpoint. We therefore consider three alternative definitions:

\begin{enumerate}
    \item The time when the total dark energy density crosses zero ($\rho_{\rm DE} = 0$);
    \item The turning point where the expansion of the universe comes to a halt ($H=0$);
    \item The big crunch time ($a=0$).
\end{enumerate}

The last choice, $a=0$, is ill-defined in numerical calculations. In practice, we terminate the integration at an epoch close to the crunch, specifically when the CMB temperature reaches $100\,\mathrm{K}$. This cutoff is also physically motivated, as the heat regulation mechanisms of habitable planets are likely to be compromised by the CMB background at such temperatures.

For each definition, we recompute the distribution of $\Omega_m$ and the probability $P(0.1<\Omega_m<0.9)$. The results are summarized in Table~\ref{tab:endpoint}. We find that while the absolute probability varies by up to $\sim 10\%$, the qualitative conclusion—that the anthropic selection naturally yields $\rho_\mathrm{m0} \sim \rho_\mathrm{DE0}$ and thereby resolves the coincidence problem—remains unchanged. This confirms that our main results are robust against reasonable variations in the anthropic window endpoint.

\begin{table}
  \centering
  \caption{{Robustness test on the choice of anthropic window endpoint.} \label{tab:endpoint}}
  \begin{tabular}{l|p{0.13\textwidth}p{0.13\textwidth}p{0.13\textwidth}}
    \hline
    \hline
     {end point} &  \multicolumn{3}{c}{{Probability of observing}}  \\
     & {$0.1<\omm<0.9$}\   & {$\rho_{\rm DE, 0} > 0$}  & {$H_0 > 0$} \\
    \hline
     {$V_{\rm eff} = 0$} & {$0.43$} & {$1$} & {$1$} \\
     {$\rho_{\rm DE} = 0$} & {$0.48$}  & {$1$} & {$1$} \\
     {$H = 0$} & {$0.31$} & {$0.66$} & {$1$} \\
     {$a = 0$} & {$0.41$} & {$0.69$} & {$0.46$} \\
    \hline
  \end{tabular}
\end{table}

Table~\ref{tab:endpoint} also lists the probabilities of observing a positive dark energy density at redshift zero ($\rho_{\rm DE,0} > 0$) and a positive Hubble constant ($H_0 > 0$). As shown in the table, these probabilities are all non-negligible across the different endpoint definitions. Thus, the ALAverse framework naturally accounts for three key features of our universe: the coincidence of $\rho_{\rm m,0}\sim \rho_{\rm DE,0}$ at the present epoch (``why now''), the positivity of the dark energy density (``why positive''), and the fact that the universe is currently expanding (``why expanding'').

\section{Comparison with observational data}\label{sec:data}

We perform our analysis using a combination of CMB, BAO, and Type Ia supernova (SNe Ia) data. For the cosmic microwave background, we adopt the Planck 2018 low-$\ell$ temperature and EE polarization likelihoods~\cite{Planck2018Like}, the Planck NPIPE high-$\ell$ CamSpec TTTEEE likelihood~\cite{Rosenberg2022}, and the Planck PR4 lensing likelihood~\cite{Carron2022}. For baryon acoustic oscillations, we include the DESI Data Release 2 (DR2) BAO measurements~\cite{DESIDR2}. For Type Ia supernovae, we use the DES-Dovekie likelihood, which is based on the reanalyzed Dark Energy Survey 5-year sample~\cite{DES-Dovekie}.

We adopt flat priors on the standard cosmological parameters $\Omega_{\mathrm b}h^2$ (baryon density), $\Omega_{\mathrm c}h^2$ (cold dark matter density), $100\theta_{\mathrm{MC}}$ (angular size of the sound horizon at recombination), $\tau_{\rm re}$ (reionization optical depth), $\ln(10^{10} A_\mathrm{s})$ (amplitude of the primordial scalar power spectrum), and $n_\mathrm{s}$ (spectral index), as well as on the dark energy parameters $\epss$ (with $|\epss|\ge 0$) and $\dom \in (-1, 1)$. The effective number of neutrino species and the sum of neutrino masses are fixed to $N_{\rm eff} = 3.044$ and $\sum m_\nu = 0.06\,\mathrm{eV}$, respectively. All parameter inferences are performed using the Cobaya~\cite{Torrado2020, Lewis2002, Neal2005, Lewis2013} and CAMB~\cite{Lewis1999, Howlett2012} codes, with CAMB modified to incorporate the new dark energy parametrization introduced in Sec.~\ref{sec:eos}.

\begin{figure}
  \centering
  \includegraphics[width=0.48\textwidth]{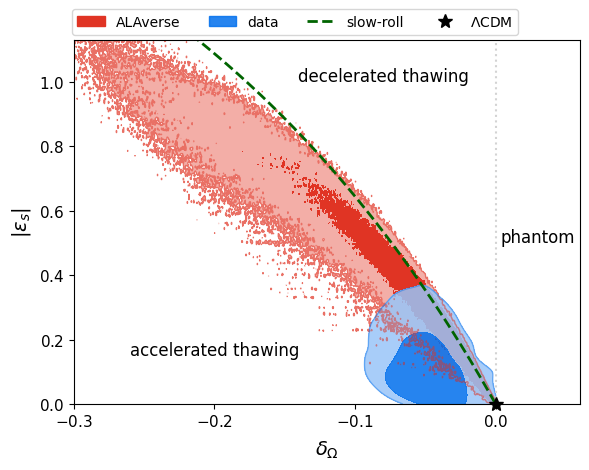}
  \caption{ALAverse solutions (with prior $\omm = 0.315\pm 0.01$ and $1+\wde< 2/3$) and observational constraints. Inner and outer contours correspond to $68.3\%$ and $95.4\%$ confidence levels, respectively. The dashed line indicates the slow-roll consistency relation from Eq.~\eqref{eq:1param_consis} evaluated at the best-fit value $\omm = 0.3156$. \label{fig:main}}
\end{figure}

Figure~\ref{fig:main} presents the key result of this work.  The marginalized constraint $\dom = -0.0498 \pm 0.0185$ indicates a $2.7\sigma$ rejection of the $\Lambda$CDM limit ($\dom = 0$) and of single-field thawing phantom models ($\dom > 0$). More general phantom scenarios, such as those allowing phantom crossing (e.g., the $w_0$-$w_a$CDM model), are not covered by this constraint and indeed provide a better fit to the data, as discussed below. 
The ALAverse solutions, sampled with $\omm = 0.315 \pm 0.01$ and $|1+\wde| < 2/3$, show an overall tendency toward accelerated thawing, i.e., $1+\wde$ rises faster than the slow-roll-limit solution~\eqref{eq:1param}. Remarkably, this same tendency is also preferred by the observational data, although the statistical significance is insufficient to fully rule out the slow-roll scenario.

Table~\ref{tab:params} lists the constraints on cosmological parameters for four models: $\Lambda$CDM, the slow-roll thawing scalar field (STSF) model defined by Eq.~\eqref{eq:1param} with $\aeq$ derived from Eq.~\eqref{eq:1param_consis}, the thawing scalar field (TSF) model given by Eq.~\eqref{eq:newparam}, and the $w_0$-$w_a$CDM model, where the dark energy EOS is parametrized as $w(a) = w_0 + w_a(1-a)$~\cite{Chevallier01,Linder03}. Below we highlight a few interesting points.

\begin{table*}
  \centering
\caption{Cosmological paramters\label{tab:params}} 
\begin{tabular}{lcccc}
  \hline
  \hline
  Parameter &  $\Lambda$CDM &  STSF, Eqs.~\eqref{eq:1param}, \eqref{eq:1param_consis} & TSF, Eq.~\eqref{eq:newparam} & $w_0$-$w_a$CDM \\
  \hline
  $\Omega_\mathrm{b} h^2$  & $0.02231\pm 0.00012$ & $0.02236\pm 0.00012$  & $0.02236\pm 0.00012$ & $0.02223\pm 0.00013$\\
  $\Omega_\mathrm{c} h^2$  & $0.11786\pm 0.00061 $ & $0.11740\pm 0.00066$ & $0.11737\pm 0.00066$ & $0.11901\pm 0.00082$ \\
  $100\theta_\mathrm{MC}$ & $1.04100\pm 0.00023$ & $1.04105\pm 0.00024$ & $1.04104\pm 0.00024$ & $1.04085\pm 0.00024$ \\  
  $\tau_\mathrm{re}$ & $0.0584\pm 0.0070$ & $0.0605\pm 0.0072$ & $0.0603\pm 0.0073$ & $0.0525\pm 0.0070$ \\
  $\ln\left(10^{10} A_\mathrm{s}\right)$  & $3.046\pm 0.014$ & $3.050\pm 0.014$ & $3.050\pm 0.014$ & $3.036\pm 0.014$ \\   
  $n_\mathrm{s}$  &  $0.9682\pm 0.0034$ & $0.9693\pm 0.0034$ & $0.9696\pm 0.0034$  & $0.9653\pm 0.0037$ \\
  $\dom$  & $0$ & $-0.0242  \pm  0.0152$ & $-0.0498\pm 0.0186$ & $-0.0570\pm 0.0184$  \\
  $|\epss|$ & $0$ & $<0.35$ (95\%CL) & $<0.29$ (95\%CL) & NA \\  
  $\epss$  & $0$ & $0.173\pm 0.109$ & $0.116_{-0.075}^{+0.101}$  & NA \\
  $w_0$  & $-1$ & NA & NA & $-0.809\pm 0.056$\\
  $w_a$ & $0$ & NA & NA & $-0.705\pm 0.217$ \\  
  \hline
  $\omm$ & $0.3038\pm 0.0036$ & $0.3099\pm 0.0053$ & $0.3157\pm 0.0059$   & $0.3132\pm 0.0053$\\  
  $\tildeomm$  & $0.3038\pm 0.0036$ & $0.3023\pm 0.0037$ & $0.2999\pm 0.0039$ & $0.2952\pm 0.0043$ \\  
  $H_0$ (km/s/Mpc) & $68.08\pm 0.27$ & $67.32\pm 0.55$ &  $66.69\pm 0.59$  & $67.32\pm 0.54$ \\
  $\sigma_8$ & $0.8061\pm 0.0057$ & $0.7976\pm 0.0077$ &  $0.7924\pm 0.0078$ & $0.8068\pm 0.0082$ \\
  \hline
  best $\chi^2$ & $12623.2$ & $12621.3$ & $12616.2$ & $12611.0$ \\
  $\Delta$AIC & $0$ & $0.1$ & $-3$ & $-8.2$ \\
  \hline
\end{tabular}
\end{table*}

First, for all models other than $\Lambda$CDM, the parameter $\tildeomm$ is more tightly constrained than $\omm$. This reflects the fact that low-redshift data measure some effective matter abundance at $z>0$. The well-known $\omm$ tension—where DESI BAO favors a lower $\omm$ than inferred from CMB~\cite{DESIDR2}—is essentially demanding $\tildeomm < \omm$, as we find in all models other than $\Lambda$CDM. Therefore, the new parameter $\tildeomm$ serves as a more robust summary statistic for low-redshift Hubble diagram measurements.

Second, the $\sigma_8$ parameter—the amplitude of linear matter density fluctuations on $8h^{-1}\,\mathrm{Mpc}$ scales—tends to be smaller in the STSF and TSF models. This may improve consistency between CMB and weak gravitational lensing measurements~\cite{Reischke2026}, although a fully self-consistent analysis remains to be performed.

Finally, model comparison based on best-fit $\chi^2$ and the Akaike Information Criterion (AIC)~\cite{Akaike1974} shows that the slow-roll thawing scalar-field (STSF) model is not preferred over $\Lambda$CDM once the penalty for the additional parameter is included. In contrast, the TSF model, which allows for accelerated thawing, is marginally preferred over $\Lambda$CDM, with $\Delta\mathrm{AIC} = -3$ (typically interpreted as weak evidence). The $w_0$-$w_a$CDM model, which permits phantom crossing ($w_{\rm DE}$ crosses $-1$), remains the statistical favorite, with $\Delta\mathrm{AIC} = -8.2$ relative to $\Lambda$CDM.

However, realizing phantom crossing in a physically well-motivated framework is nontrivial; it often requires exotic field theories with negative kinetic terms or modifications of gravity. By contrast, the TSF behavior considered here arises naturally within the ALAverse, driven by the interplay between the negative cosmological constant and the axion dynamics. Moreover, as we have argued, the ALAverse simultaneously addresses the fine-tuning and coincidence problems of dark energy—a feature that purely phenomenological parametrizations such as $w_0$-$w_a$CDM cannot offer. Thus, although TSF is not the statistical best fit among the models considered, its theoretical foundation and its ability to resolve conceptual problems make it a compelling candidate for the dark energy sector.

\section{Discussion and conclusions}\label{sec:conc}

In this work, we have introduced the ALAverse, a framework that naturally connects the string axiverse with a negative cosmological constant through an observation-time-weighted anthropic selection. The combination of a positive axion potential with a negative cosmological constant offers a key advantage: the coincidence problem can be resolved without fine-tuning, as the anthropic weighting automatically favors cosmologies with observation times comparable to the matter-dominated epoch.  We emphasize that the anthropic argument employed here does not claim to explain why the bare cosmological constant is not of order $\mathrm{TeV}^4$ or higher—this remains an open question in the string landscape, as it does in all anthropic approaches to the cosmological constant problem. Rather, we assume that the landscape provides a broad prior over the effective dark energy density, and that the anthropic weighting then selects those universes in which this density is sufficiently small to permit structure formation. Within this standard framework, the ALAverse offers a concrete and falsifiable realization in which a negative cosmological constant is effectively compensated by a positive axion component, and the observation-time-weighted anthropic principle naturally resolves the coincidence problem.

It is worth clarifying the logical structure of our predictions. The anthropic argument provides an unconditional prediction that observing  $\rho_{\mathrm m,0}\sim\rho_{\mathrm{DE},0}$ is not a finely tuned accident. Once $\omm$ is limited to the observed range ($\omm \simeq 0.315$), the model then predicts a specific EOS evolution, namely accelerated thawing. This conditional prediction is falsifiable: future high-precision measurements of the Hubble diagram at $\omm \simeq 0.315$ can confirm or rule it out, independently of whether the same signature persists at other values of $\omm$ that are not realized in our universe.

 Conditioned on the observed value $\omm \simeq 0.315$, we have shown that the ALAverse statistically favors an accelerated-thawing dynamics, leading to a new parametrization of dark energy that extends beyond the standard slow-roll scenario. When confronted with recent CMB, BAO, and SNIa data, our model exhibits a mild preference over $\Lambda$CDM and provides a physically motivated alternative to purely phenomenological parametrizations such as $w_0$-$w_a$CDM. Although the statistical evidence for accelerated thawing—the key signature of the ALAverse—remains moderate, the ALAverse+TSF parametrization framework renders the scenario falsifiable with future precision measurements of the late-universe Hubble diagram. To the best of our knowledge, this is the first string-theory-inspired anthropic model that is testable and falsifiable by precision cosmological data.



\section{Acknowledgements}

This work is supported by the National Natural Science Foundation of China (NSFC) under its Key Program (Grant No. 12533002). We also note, with amusement, that ``ALA'' coincidentally means ``our'' in the dialect of the author's hometown, rendering ``ALAverse'' a pleasantly fitting name for our universe.

%


\begin{thebibliography}{45}%
\makeatletter
\providecommand \@ifxundefined [1]{%
 \@ifx{#1\undefined}
}%
\providecommand \@ifnum [1]{%
 \ifnum #1\expandafter \@firstoftwo
 \else \expandafter \@secondoftwo
 \fi
}%
\providecommand \@ifx [1]{%
 \ifx #1\expandafter \@firstoftwo
 \else \expandafter \@secondoftwo
 \fi
}%
\providecommand \natexlab [1]{#1}%
\providecommand \enquote  [1]{``#1''}%
\providecommand \bibnamefont  [1]{#1}%
\providecommand \bibfnamefont [1]{#1}%
\providecommand \citenamefont [1]{#1}%
\providecommand \href@noop [0]{\@secondoftwo}%
\providecommand \href [0]{\begingroup \@sanitize@url \@href}%
\providecommand \@href[1]{\@@startlink{#1}\@@href}%
\providecommand \@@href[1]{\endgroup#1\@@endlink}%
\providecommand \@sanitize@url [0]{\catcode `\\12\catcode `\$12\catcode
  `\&12\catcode `\#12\catcode `\^12\catcode `\_12\catcode `\%12\relax}%
\providecommand \@@startlink[1]{}%
\providecommand \@@endlink[0]{}%
\providecommand \url  [0]{\begingroup\@sanitize@url \@url }%
\providecommand \@url [1]{\endgroup\@href {#1}{\urlprefix }}%
\providecommand \urlprefix  [0]{URL }%
\providecommand \Eprint [0]{\href }%
\providecommand \doibase [0]{http://dx.doi.org/}%
\providecommand \selectlanguage [0]{\@gobble}%
\providecommand \bibinfo  [0]{\@secondoftwo}%
\providecommand \bibfield  [0]{\@secondoftwo}%
\providecommand \translation [1]{[#1]}%
\providecommand \BibitemOpen [0]{}%
\providecommand \bibitemStop [0]{}%
\providecommand \bibitemNoStop [0]{.\EOS\space}%
\providecommand \EOS [0]{\spacefactor3000\relax}%
\providecommand \BibitemShut  [1]{\csname bibitem#1\endcsname}%
\let\auto@bib@innerbib\@empty
\bibitem [{\citenamefont {{Riess}}\ \emph {et~al.}(1998)\citenamefont
  {{Riess}}, \citenamefont {{Filippenko}}, \citenamefont {{Challis}},
  \citenamefont {{Clocchiatti}}, \citenamefont {{Diercks}}, \citenamefont
  {{Garnavich}}, \citenamefont {{Gilliland}}, \citenamefont {{Hogan}},
  \citenamefont {{Jha}}, \citenamefont {{Kirshner}}, \citenamefont
  {{Leibundgut}}, \citenamefont {{Phillips}}, \citenamefont {{Reiss}},
  \citenamefont {{Schmidt}}, \citenamefont {{Schommer}}, \citenamefont
  {{Smith}}, \citenamefont {{Spyromilio}}, \citenamefont {{Stubbs}},
  \citenamefont {{Suntzeff}},\ and\ \citenamefont {{Tonry}}}]{Riess1998}%
  \BibitemOpen
  \bibfield  {author} {\bibinfo {author} {\bibfnamefont {A.~G.}\ \bibnamefont
  {{Riess}}}, \bibinfo {author} {\bibfnamefont {A.~V.}\ \bibnamefont
  {{Filippenko}}}, \bibinfo {author} {\bibfnamefont {P.}~\bibnamefont
  {{Challis}}}, \bibinfo {author} {\bibfnamefont {A.}~\bibnamefont
  {{Clocchiatti}}}, \bibinfo {author} {\bibfnamefont {A.}~\bibnamefont
  {{Diercks}}}, \bibinfo {author} {\bibfnamefont {P.~M.}\ \bibnamefont
  {{Garnavich}}}, \bibinfo {author} {\bibfnamefont {R.~L.}\ \bibnamefont
  {{Gilliland}}}, \bibinfo {author} {\bibfnamefont {C.~J.}\ \bibnamefont
  {{Hogan}}}, \bibinfo {author} {\bibfnamefont {S.}~\bibnamefont {{Jha}}},
  \bibinfo {author} {\bibfnamefont {R.~P.}\ \bibnamefont {{Kirshner}}},
  \bibinfo {author} {\bibfnamefont {B.}~\bibnamefont {{Leibundgut}}}, \bibinfo
  {author} {\bibfnamefont {M.~M.}\ \bibnamefont {{Phillips}}}, \bibinfo
  {author} {\bibfnamefont {D.}~\bibnamefont {{Reiss}}}, \bibinfo {author}
  {\bibfnamefont {B.~P.}\ \bibnamefont {{Schmidt}}}, \bibinfo {author}
  {\bibfnamefont {R.~A.}\ \bibnamefont {{Schommer}}}, \bibinfo {author}
  {\bibfnamefont {R.~C.}\ \bibnamefont {{Smith}}}, \bibinfo {author}
  {\bibfnamefont {J.}~\bibnamefont {{Spyromilio}}}, \bibinfo {author}
  {\bibfnamefont {C.}~\bibnamefont {{Stubbs}}}, \bibinfo {author}
  {\bibfnamefont {N.~B.}\ \bibnamefont {{Suntzeff}}}, \ and\ \bibinfo {author}
  {\bibfnamefont {J.}~\bibnamefont {{Tonry}}},\ }\href {\doibase
  10.1086/300499} {\bibfield  {journal} {\bibinfo  {journal} {\aj}\ }\textbf
  {\bibinfo {volume} {116}},\ \bibinfo {pages} {1009} (\bibinfo {year}
  {1998})},\ \Eprint {http://arxiv.org/abs/astro-ph/9805201}
  {arXiv:astro-ph/9805201 [astro-ph]} \BibitemShut {NoStop}%
\bibitem [{\citenamefont {{Perlmutter}}\ \emph {et~al.}(1999)\citenamefont
  {{Perlmutter}}, \citenamefont {{Aldering}}, \citenamefont {{Goldhaber}},
  \citenamefont {{Knop}}, \citenamefont {{Nugent}}, \citenamefont {{Castro}},
  \citenamefont {{Deustua}}, \citenamefont {{Fabbro}}, \citenamefont
  {{Goobar}}, \citenamefont {{Groom}}, \citenamefont {{Hook}}, \citenamefont
  {{Kim}}, \citenamefont {{Kim}}, \citenamefont {{Lee}}, \citenamefont
  {{Nunes}}, \citenamefont {{Pain}}, \citenamefont {{Pennypacker}},
  \citenamefont {{Quimby}}, \citenamefont {{Lidman}}, \citenamefont {{Ellis}},
  \citenamefont {{Irwin}}, \citenamefont {{McMahon}}, \citenamefont
  {{Ruiz-Lapuente}}, \citenamefont {{Walton}}, \citenamefont {{Schaefer}},
  \citenamefont {{Boyle}}, \citenamefont {{Filippenko}}, \citenamefont
  {{Matheson}}, \citenamefont {{Fruchter}}, \citenamefont {{Panagia}},
  \citenamefont {{Newberg}}, \citenamefont {{Couch}},\ and\ \citenamefont
  {{Project}}}]{Perlmutter1999}%
  \BibitemOpen
  \bibfield  {author} {\bibinfo {author} {\bibfnamefont {S.}~\bibnamefont
  {{Perlmutter}}}, \bibinfo {author} {\bibfnamefont {G.}~\bibnamefont
  {{Aldering}}}, \bibinfo {author} {\bibfnamefont {G.}~\bibnamefont
  {{Goldhaber}}}, \bibinfo {author} {\bibfnamefont {R.~A.}\ \bibnamefont
  {{Knop}}}, \bibinfo {author} {\bibfnamefont {P.}~\bibnamefont {{Nugent}}},
  \bibinfo {author} {\bibfnamefont {P.~G.}\ \bibnamefont {{Castro}}}, \bibinfo
  {author} {\bibfnamefont {S.}~\bibnamefont {{Deustua}}}, \bibinfo {author}
  {\bibfnamefont {S.}~\bibnamefont {{Fabbro}}}, \bibinfo {author}
  {\bibfnamefont {A.}~\bibnamefont {{Goobar}}}, \bibinfo {author}
  {\bibfnamefont {D.~E.}\ \bibnamefont {{Groom}}}, \bibinfo {author}
  {\bibfnamefont {I.~M.}\ \bibnamefont {{Hook}}}, \bibinfo {author}
  {\bibfnamefont {A.~G.}\ \bibnamefont {{Kim}}}, \bibinfo {author}
  {\bibfnamefont {M.~Y.}\ \bibnamefont {{Kim}}}, \bibinfo {author}
  {\bibfnamefont {J.~C.}\ \bibnamefont {{Lee}}}, \bibinfo {author}
  {\bibfnamefont {N.~J.}\ \bibnamefont {{Nunes}}}, \bibinfo {author}
  {\bibfnamefont {R.}~\bibnamefont {{Pain}}}, \bibinfo {author} {\bibfnamefont
  {C.~R.}\ \bibnamefont {{Pennypacker}}}, \bibinfo {author} {\bibfnamefont
  {R.}~\bibnamefont {{Quimby}}}, \bibinfo {author} {\bibfnamefont
  {C.}~\bibnamefont {{Lidman}}}, \bibinfo {author} {\bibfnamefont {R.~S.}\
  \bibnamefont {{Ellis}}}, \bibinfo {author} {\bibfnamefont {M.}~\bibnamefont
  {{Irwin}}}, \bibinfo {author} {\bibfnamefont {R.~G.}\ \bibnamefont
  {{McMahon}}}, \bibinfo {author} {\bibfnamefont {P.}~\bibnamefont
  {{Ruiz-Lapuente}}}, \bibinfo {author} {\bibfnamefont {N.}~\bibnamefont
  {{Walton}}}, \bibinfo {author} {\bibfnamefont {B.}~\bibnamefont
  {{Schaefer}}}, \bibinfo {author} {\bibfnamefont {B.~J.}\ \bibnamefont
  {{Boyle}}}, \bibinfo {author} {\bibfnamefont {A.~V.}\ \bibnamefont
  {{Filippenko}}}, \bibinfo {author} {\bibfnamefont {T.}~\bibnamefont
  {{Matheson}}}, \bibinfo {author} {\bibfnamefont {A.~S.}\ \bibnamefont
  {{Fruchter}}}, \bibinfo {author} {\bibfnamefont {N.}~\bibnamefont
  {{Panagia}}}, \bibinfo {author} {\bibfnamefont {H.~J.~M.}\ \bibnamefont
  {{Newberg}}}, \bibinfo {author} {\bibfnamefont {W.~J.}\ \bibnamefont
  {{Couch}}}, \ and\ \bibinfo {author} {\bibfnamefont {T.~S.~C.}\ \bibnamefont
  {{Project}}},\ }\href {\doibase 10.1086/307221} {\bibfield  {journal}
  {\bibinfo  {journal} {\apj}\ }\textbf {\bibinfo {volume} {517}},\ \bibinfo
  {pages} {565} (\bibinfo {year} {1999})},\ \Eprint
  {http://arxiv.org/abs/astro-ph/9812133} {arXiv:astro-ph/9812133 [astro-ph]}
  \BibitemShut {NoStop}%
\bibitem [{\citenamefont {{Aghanim}}\ \emph
  {et~al.}(2020{\natexlab{a}})\citenamefont {{Aghanim}}, \citenamefont
  {{Akrami}}, \citenamefont {{Ashdown}} \emph {et~al.}}]{Planck2018Params}%
  \BibitemOpen
  \bibfield  {author} {\bibinfo {author} {\bibfnamefont {N.}~\bibnamefont
  {{Aghanim}}}, \bibinfo {author} {\bibfnamefont {Y.}~\bibnamefont {{Akrami}}},
  \bibinfo {author} {\bibfnamefont {M.}~\bibnamefont {{Ashdown}}},  \emph
  {et~al.},\ }\href {\doibase 10.1051/0004-6361/201833910} {\bibfield
  {journal} {\bibinfo  {journal} {\aap}\ }\textbf {\bibinfo {volume} {641}},\
  \bibinfo {eid} {A6} (\bibinfo {year} {2020}{\natexlab{a}})},\ \Eprint
  {http://arxiv.org/abs/1807.06209} {arXiv:1807.06209 [astro-ph.CO]}
  \BibitemShut {NoStop}%
\bibitem [{\citenamefont {{Weinberg}}(1989)}]{Weinberg1989}%
  \BibitemOpen
  \bibfield  {author} {\bibinfo {author} {\bibfnamefont {S.}~\bibnamefont
  {{Weinberg}}},\ }\href {\doibase 10.1103/RevModPhys.61.1} {\bibfield
  {journal} {\bibinfo  {journal} {Reviews of Modern Physics}\ }\textbf
  {\bibinfo {volume} {61}},\ \bibinfo {pages} {1} (\bibinfo {year}
  {1989})}\BibitemShut {NoStop}%
\bibitem [{\citenamefont {{Kachru}}\ \emph {et~al.}(2003)\citenamefont
  {{Kachru}}, \citenamefont {{Kallosh}}, \citenamefont {{Linde}},\ and\
  \citenamefont {{Trivedi}}}]{KKLT}%
  \BibitemOpen
  \bibfield  {author} {\bibinfo {author} {\bibfnamefont {S.}~\bibnamefont
  {{Kachru}}}, \bibinfo {author} {\bibfnamefont {R.}~\bibnamefont {{Kallosh}}},
  \bibinfo {author} {\bibfnamefont {A.}~\bibnamefont {{Linde}}}, \ and\
  \bibinfo {author} {\bibfnamefont {S.~P.}\ \bibnamefont {{Trivedi}}},\ }\href
  {\doibase 10.1103/PhysRevD.68.046005} {\bibfield  {journal} {\bibinfo
  {journal} {\prd}\ }\textbf {\bibinfo {volume} {68}},\ \bibinfo {eid} {046005}
  (\bibinfo {year} {2003})},\ \Eprint {http://arxiv.org/abs/hep-th/0301240}
  {arXiv:hep-th/0301240 [hep-th]} \BibitemShut {NoStop}%
\bibitem [{\citenamefont {{Carter}}(1974)}]{Carter1974}%
  \BibitemOpen
  \bibfield  {author} {\bibinfo {author} {\bibfnamefont {B.}~\bibnamefont
  {{Carter}}},\ }in\ \href@noop {} {\emph {\bibinfo {booktitle} {Confrontation
  of Cosmological Theories with Observational Data}}},\ \bibinfo {series} {IAU
  Symposium}, Vol.~\bibinfo {volume} {63},\ \bibinfo {editor} {edited by\
  \bibinfo {editor} {\bibfnamefont {M.~S.}\ \bibnamefont {{Longair}}}}\
  (\bibinfo {year} {1974})\ pp.\ \bibinfo {pages} {291--298}\BibitemShut
  {NoStop}%
\bibitem [{\citenamefont {{Weinberg}}(1987)}]{Weinberg1987}%
  \BibitemOpen
  \bibfield  {author} {\bibinfo {author} {\bibfnamefont {S.}~\bibnamefont
  {{Weinberg}}},\ }\href {\doibase 10.1103/PhysRevLett.59.2607} {\bibfield
  {journal} {\bibinfo  {journal} {\prl}\ }\textbf {\bibinfo {volume} {59}},\
  \bibinfo {pages} {2607} (\bibinfo {year} {1987})}\BibitemShut {NoStop}%
\bibitem [{\citenamefont {{Zlatev}}\ \emph {et~al.}(1999)\citenamefont
  {{Zlatev}}, \citenamefont {{Wang}},\ and\ \citenamefont
  {{Steinhardt}}}]{Zlatev1999}%
  \BibitemOpen
  \bibfield  {author} {\bibinfo {author} {\bibfnamefont {I.}~\bibnamefont
  {{Zlatev}}}, \bibinfo {author} {\bibfnamefont {L.}~\bibnamefont {{Wang}}}, \
  and\ \bibinfo {author} {\bibfnamefont {P.~J.}\ \bibnamefont {{Steinhardt}}},\
  }\href {\doibase 10.1103/PhysRevLett.82.896} {\bibfield  {journal} {\bibinfo
  {journal} {\prl}\ }\textbf {\bibinfo {volume} {82}},\ \bibinfo {pages} {896}
  (\bibinfo {year} {1999})},\ \Eprint {http://arxiv.org/abs/astro-ph/9807002}
  {arXiv:astro-ph/9807002 [astro-ph]} \BibitemShut {NoStop}%
\bibitem [{\citenamefont {{Arvanitaki}}\ \emph {et~al.}(2010)\citenamefont
  {{Arvanitaki}}, \citenamefont {{Dimopoulos}}, \citenamefont {{Dubovsky}},
  \citenamefont {{Kaloper}},\ and\ \citenamefont
  {{March-Russell}}}]{Arvanitaki2010}%
  \BibitemOpen
  \bibfield  {author} {\bibinfo {author} {\bibfnamefont {A.}~\bibnamefont
  {{Arvanitaki}}}, \bibinfo {author} {\bibfnamefont {S.}~\bibnamefont
  {{Dimopoulos}}}, \bibinfo {author} {\bibfnamefont {S.}~\bibnamefont
  {{Dubovsky}}}, \bibinfo {author} {\bibfnamefont {N.}~\bibnamefont
  {{Kaloper}}}, \ and\ \bibinfo {author} {\bibfnamefont {J.}~\bibnamefont
  {{March-Russell}}},\ }\href {\doibase 10.1103/PhysRevD.81.123530} {\bibfield
  {journal} {\bibinfo  {journal} {\prd}\ }\textbf {\bibinfo {volume} {81}},\
  \bibinfo {eid} {123530} (\bibinfo {year} {2010})},\ \Eprint
  {http://arxiv.org/abs/0905.4720} {arXiv:0905.4720 [hep-th]} \BibitemShut
  {NoStop}%
\bibitem [{\citenamefont {{Freese}}\ \emph {et~al.}(1990)\citenamefont
  {{Freese}}, \citenamefont {{Frieman}},\ and\ \citenamefont
  {{Olinto}}}]{Freese1990}%
  \BibitemOpen
  \bibfield  {author} {\bibinfo {author} {\bibfnamefont {K.}~\bibnamefont
  {{Freese}}}, \bibinfo {author} {\bibfnamefont {J.~A.}\ \bibnamefont
  {{Frieman}}}, \ and\ \bibinfo {author} {\bibfnamefont {A.~V.}\ \bibnamefont
  {{Olinto}}},\ }\href {\doibase 10.1103/PhysRevLett.65.3233} {\bibfield
  {journal} {\bibinfo  {journal} {Phys. Rev. Lett.}\ }\textbf {\bibinfo
  {volume} {65}},\ \bibinfo {pages} {3233} (\bibinfo {year}
  {1990})}\BibitemShut {NoStop}%
\bibitem [{\citenamefont {{Odintsov}}\ and\ \citenamefont
  {{Oikonomou}}(2019)}]{Odintsov2019}%
  \BibitemOpen
  \bibfield  {author} {\bibinfo {author} {\bibfnamefont {S.~D.}\ \bibnamefont
  {{Odintsov}}}\ and\ \bibinfo {author} {\bibfnamefont {V.~K.}\ \bibnamefont
  {{Oikonomou}}},\ }\href {\doibase 10.1103/PhysRevD.99.104070} {\bibfield
  {journal} {\bibinfo  {journal} {\prd}\ }\textbf {\bibinfo {volume} {99}},\
  \bibinfo {eid} {104070} (\bibinfo {year} {2019})},\ \Eprint
  {http://arxiv.org/abs/1905.03496} {arXiv:1905.03496 [gr-qc]} \BibitemShut
  {NoStop}%
\bibitem [{\citenamefont {{Danielsson}}\ and\ \citenamefont
  {{Riet}}(2018)}]{Danielsson2018}%
  \BibitemOpen
  \bibfield  {author} {\bibinfo {author} {\bibfnamefont {U.~H.}\ \bibnamefont
  {{Danielsson}}}\ and\ \bibinfo {author} {\bibfnamefont {T.~V.}\ \bibnamefont
  {{Riet}}},\ }\href {\doibase 10.1142/S0218271818300070} {\bibfield  {journal}
  {\bibinfo  {journal} {International Journal of Modern Physics D}\ }\textbf
  {\bibinfo {volume} {27}},\ \bibinfo {eid} {1830007-298} (\bibinfo {year}
  {2018})},\ \Eprint {http://arxiv.org/abs/1804.01120} {arXiv:1804.01120
  [hep-th]} \BibitemShut {NoStop}%
\bibitem [{\citenamefont {{Murai}}\ and\ \citenamefont
  {{Takahashi}}(2025)}]{Murai2025}%
  \BibitemOpen
  \bibfield  {author} {\bibinfo {author} {\bibfnamefont {K.}~\bibnamefont
  {{Murai}}}\ and\ \bibinfo {author} {\bibfnamefont {F.}~\bibnamefont
  {{Takahashi}}},\ }\href {\doibase 10.1103/kwhj-vl35} {\bibfield  {journal}
  {\bibinfo  {journal} {\prd}\ }\textbf {\bibinfo {volume} {112}},\ \bibinfo
  {eid} {103501} (\bibinfo {year} {2025})},\ \Eprint
  {http://arxiv.org/abs/2504.12852} {arXiv:2504.12852 [hep-ph]} \BibitemShut
  {NoStop}%
\bibitem [{\citenamefont {{Luu}}\ \emph
  {et~al.}(2025{\natexlab{a}})\citenamefont {{Luu}}, \citenamefont {{Qiu}},\
  and\ \citenamefont {{Tye}}}]{Luu2025a}%
  \BibitemOpen
  \bibfield  {author} {\bibinfo {author} {\bibfnamefont {H.~N.}\ \bibnamefont
  {{Luu}}}, \bibinfo {author} {\bibfnamefont {Y.-C.}\ \bibnamefont {{Qiu}}}, \
  and\ \bibinfo {author} {\bibfnamefont {S.-H.~H.}\ \bibnamefont {{Tye}}},\
  }\href {\doibase 10.1103/3mpg-24d2} {\bibfield  {journal} {\bibinfo
  {journal} {\prd}\ }\textbf {\bibinfo {volume} {112}},\ \bibinfo {eid}
  {023524} (\bibinfo {year} {2025}{\natexlab{a}})},\ \Eprint
  {http://arxiv.org/abs/2503.18120} {arXiv:2503.18120 [hep-ph]} \BibitemShut
  {NoStop}%
\bibitem [{\citenamefont {{Starkman}}\ and\ \citenamefont
  {{Trotta}}(2006)}]{Starkman2006}%
  \BibitemOpen
  \bibfield  {author} {\bibinfo {author} {\bibfnamefont {G.~D.}\ \bibnamefont
  {{Starkman}}}\ and\ \bibinfo {author} {\bibfnamefont {R.}~\bibnamefont
  {{Trotta}}},\ }\href {\doibase 10.1103/PhysRevLett.97.201301} {\bibfield
  {journal} {\bibinfo  {journal} {\prl}\ }\textbf {\bibinfo {volume} {97}},\
  \bibinfo {eid} {201301} (\bibinfo {year} {2006})},\ \Eprint
  {http://arxiv.org/abs/astro-ph/0607227} {arXiv:astro-ph/0607227 [astro-ph]}
  \BibitemShut {NoStop}%
\bibitem [{\citenamefont {{Sudoh}}\ \emph {et~al.}(2017)\citenamefont
  {{Sudoh}}, \citenamefont {{Totani}}, \citenamefont {{Makiya}},\ and\
  \citenamefont {{Nagashima}}}]{Sudoh2017}%
  \BibitemOpen
  \bibfield  {author} {\bibinfo {author} {\bibfnamefont {T.}~\bibnamefont
  {{Sudoh}}}, \bibinfo {author} {\bibfnamefont {T.}~\bibnamefont {{Totani}}},
  \bibinfo {author} {\bibfnamefont {R.}~\bibnamefont {{Makiya}}}, \ and\
  \bibinfo {author} {\bibfnamefont {M.}~\bibnamefont {{Nagashima}}},\ }\href
  {\doibase 10.1093/mnras/stw2401} {\bibfield  {journal} {\bibinfo  {journal}
  {\mnras}\ }\textbf {\bibinfo {volume} {464}},\ \bibinfo {pages} {1563}
  (\bibinfo {year} {2017})},\ \Eprint {http://arxiv.org/abs/1607.00180}
  {arXiv:1607.00180 [astro-ph.CO]} \BibitemShut {NoStop}%
\bibitem [{\citenamefont {{Sorini}}\ \emph {et~al.}(2024)\citenamefont
  {{Sorini}}, \citenamefont {{Peacock}},\ and\ \citenamefont
  {{Lombriser}}}]{Sorini2024}%
  \BibitemOpen
  \bibfield  {author} {\bibinfo {author} {\bibfnamefont {D.}~\bibnamefont
  {{Sorini}}}, \bibinfo {author} {\bibfnamefont {J.~A.}\ \bibnamefont
  {{Peacock}}}, \ and\ \bibinfo {author} {\bibfnamefont {L.}~\bibnamefont
  {{Lombriser}}},\ }\href {\doibase 10.1093/mnras/stae2236} {\bibfield
  {journal} {\bibinfo  {journal} {\mnras}\ }\textbf {\bibinfo {volume} {535}},\
  \bibinfo {pages} {1449} (\bibinfo {year} {2024})},\ \Eprint
  {http://arxiv.org/abs/2411.07301} {arXiv:2411.07301 [astro-ph.CO]}
  \BibitemShut {NoStop}%
\bibitem [{\citenamefont {{Luu}}\ \emph
  {et~al.}(2025{\natexlab{b}})\citenamefont {{Luu}}, \citenamefont {{Qiu}},\
  and\ \citenamefont {{Tye}}}]{Luu2025b}%
  \BibitemOpen
  \bibfield  {author} {\bibinfo {author} {\bibfnamefont {H.~N.}\ \bibnamefont
  {{Luu}}}, \bibinfo {author} {\bibfnamefont {Y.-C.}\ \bibnamefont {{Qiu}}}, \
  and\ \bibinfo {author} {\bibfnamefont {S.-H.~H.}\ \bibnamefont {{Tye}}},\
  }\href {\doibase 10.1088/1475-7516/2025/09/055} {\bibfield  {journal}
  {\bibinfo  {journal} {\jcap}\ }\textbf {\bibinfo {volume} {2025}},\ \bibinfo
  {eid} {055} (\bibinfo {year} {2025}{\natexlab{b}})},\ \Eprint
  {http://arxiv.org/abs/2506.24011} {arXiv:2506.24011 [hep-ph]} \BibitemShut
  {NoStop}%
\bibitem [{\citenamefont {{Efstathiou}}(1995)}]{Efstathiou1995}%
  \BibitemOpen
  \bibfield  {author} {\bibinfo {author} {\bibfnamefont {G.}~\bibnamefont
  {{Efstathiou}}},\ }\href {\doibase 10.1093/mnras/274.1.L73} {\bibfield
  {journal} {\bibinfo  {journal} {\mnras}\ }\textbf {\bibinfo {volume} {274}},\
  \bibinfo {pages} {L73} (\bibinfo {year} {1995})}\BibitemShut {NoStop}%
\bibitem [{\citenamefont {{Garriga}}\ \emph {et~al.}(1999)\citenamefont
  {{Garriga}}, \citenamefont {{Livio}},\ and\ \citenamefont
  {{Vilenkin}}}]{Garriga1999}%
  \BibitemOpen
  \bibfield  {author} {\bibinfo {author} {\bibfnamefont {J.}~\bibnamefont
  {{Garriga}}}, \bibinfo {author} {\bibfnamefont {M.}~\bibnamefont {{Livio}}},
  \ and\ \bibinfo {author} {\bibfnamefont {A.}~\bibnamefont {{Vilenkin}}},\
  }\href {\doibase 10.1103/PhysRevD.61.023503} {\bibfield  {journal} {\bibinfo
  {journal} {\prd}\ }\textbf {\bibinfo {volume} {61}},\ \bibinfo {eid} {023503}
  (\bibinfo {year} {1999})},\ \Eprint {http://arxiv.org/abs/astro-ph/9906210}
  {arXiv:astro-ph/9906210 [astro-ph]} \BibitemShut {NoStop}%
\bibitem [{\citenamefont {{Peacock}}(2007)}]{Peacock2007}%
  \BibitemOpen
  \bibfield  {author} {\bibinfo {author} {\bibfnamefont {J.~A.}\ \bibnamefont
  {{Peacock}}},\ }\href {\doibase 10.1111/j.1365-2966.2007.11978.x} {\bibfield
  {journal} {\bibinfo  {journal} {\mnras}\ }\textbf {\bibinfo {volume} {379}},\
  \bibinfo {pages} {1067} (\bibinfo {year} {2007})},\ \Eprint
  {http://arxiv.org/abs/0705.0898} {arXiv:0705.0898 [astro-ph]} \BibitemShut
  {NoStop}%
\bibitem [{\citenamefont {{Bousso}}\ and\ \citenamefont
  {{Leichenauer}}(2010)}]{Bousso2010}%
  \BibitemOpen
  \bibfield  {author} {\bibinfo {author} {\bibfnamefont {R.}~\bibnamefont
  {{Bousso}}}\ and\ \bibinfo {author} {\bibfnamefont {S.}~\bibnamefont
  {{Leichenauer}}},\ }\href {\doibase 10.1103/PhysRevD.81.063524} {\bibfield
  {journal} {\bibinfo  {journal} {\prd}\ }\textbf {\bibinfo {volume} {81}},\
  \bibinfo {eid} {063524} (\bibinfo {year} {2010})},\ \Eprint
  {http://arxiv.org/abs/0907.4917} {arXiv:0907.4917 [hep-th]} \BibitemShut
  {NoStop}%
\bibitem [{\citenamefont {{Barnes}}\ \emph {et~al.}(2018)\citenamefont
  {{Barnes}}, \citenamefont {{Elahi}}, \citenamefont {{Salcido}}, \citenamefont
  {{Bower}}, \citenamefont {{Lewis}}, \citenamefont {{Theuns}}, \citenamefont
  {{Schaller}}, \citenamefont {{Crain}},\ and\ \citenamefont
  {{Schaye}}}]{Barnes2018}%
  \BibitemOpen
  \bibfield  {author} {\bibinfo {author} {\bibfnamefont {L.~A.}\ \bibnamefont
  {{Barnes}}}, \bibinfo {author} {\bibfnamefont {P.~J.}\ \bibnamefont
  {{Elahi}}}, \bibinfo {author} {\bibfnamefont {J.}~\bibnamefont {{Salcido}}},
  \bibinfo {author} {\bibfnamefont {R.~G.}\ \bibnamefont {{Bower}}}, \bibinfo
  {author} {\bibfnamefont {G.~F.}\ \bibnamefont {{Lewis}}}, \bibinfo {author}
  {\bibfnamefont {T.}~\bibnamefont {{Theuns}}}, \bibinfo {author}
  {\bibfnamefont {M.}~\bibnamefont {{Schaller}}}, \bibinfo {author}
  {\bibfnamefont {R.~A.}\ \bibnamefont {{Crain}}}, \ and\ \bibinfo {author}
  {\bibfnamefont {J.}~\bibnamefont {{Schaye}}},\ }\href {\doibase
  10.1093/mnras/sty846} {\bibfield  {journal} {\bibinfo  {journal} {\mnras}\
  }\textbf {\bibinfo {volume} {477}},\ \bibinfo {pages} {3727} (\bibinfo {year}
  {2018})},\ \Eprint {http://arxiv.org/abs/1801.08781} {arXiv:1801.08781
  [astro-ph.CO]} \BibitemShut {NoStop}%
\bibitem [{\citenamefont {{Salcido}}\ \emph {et~al.}(2018)\citenamefont
  {{Salcido}}, \citenamefont {{Bower}}, \citenamefont {{Barnes}}, \citenamefont
  {{Lewis}}, \citenamefont {{Elahi}}, \citenamefont {{Theuns}}, \citenamefont
  {{Schaller}}, \citenamefont {{Crain}},\ and\ \citenamefont
  {{Schaye}}}]{Salcido2018}%
  \BibitemOpen
  \bibfield  {author} {\bibinfo {author} {\bibfnamefont {J.}~\bibnamefont
  {{Salcido}}}, \bibinfo {author} {\bibfnamefont {R.~G.}\ \bibnamefont
  {{Bower}}}, \bibinfo {author} {\bibfnamefont {L.~A.}\ \bibnamefont
  {{Barnes}}}, \bibinfo {author} {\bibfnamefont {G.~F.}\ \bibnamefont
  {{Lewis}}}, \bibinfo {author} {\bibfnamefont {P.~J.}\ \bibnamefont
  {{Elahi}}}, \bibinfo {author} {\bibfnamefont {T.}~\bibnamefont {{Theuns}}},
  \bibinfo {author} {\bibfnamefont {M.}~\bibnamefont {{Schaller}}}, \bibinfo
  {author} {\bibfnamefont {R.~A.}\ \bibnamefont {{Crain}}}, \ and\ \bibinfo
  {author} {\bibfnamefont {J.}~\bibnamefont {{Schaye}}},\ }\href {\doibase
  10.1093/mnras/sty879} {\bibfield  {journal} {\bibinfo  {journal} {\mnras}\
  }\textbf {\bibinfo {volume} {477}},\ \bibinfo {pages} {3744} (\bibinfo {year}
  {2018})},\ \Eprint {http://arxiv.org/abs/1710.06861} {arXiv:1710.06861
  [astro-ph.CO]} \BibitemShut {NoStop}%
\bibitem [{\citenamefont {{Oh}}\ \emph {et~al.}(2022)\citenamefont {{Oh}},
  \citenamefont {{Peacock}}, \citenamefont {{Khochfar}},\ and\ \citenamefont
  {{Smith}}}]{Oh2022}%
  \BibitemOpen
  \bibfield  {author} {\bibinfo {author} {\bibfnamefont {B.~K.}\ \bibnamefont
  {{Oh}}}, \bibinfo {author} {\bibfnamefont {J.~A.}\ \bibnamefont {{Peacock}}},
  \bibinfo {author} {\bibfnamefont {S.}~\bibnamefont {{Khochfar}}}, \ and\
  \bibinfo {author} {\bibfnamefont {B.~D.}\ \bibnamefont {{Smith}}},\ }\href
  {\doibase 10.1093/mnras/stac2669} {\bibfield  {journal} {\bibinfo  {journal}
  {\mnras}\ }\textbf {\bibinfo {volume} {517}},\ \bibinfo {pages} {59}
  (\bibinfo {year} {2022})},\ \Eprint {http://arxiv.org/abs/2209.08783}
  {arXiv:2209.08783 [astro-ph.CO]} \BibitemShut {NoStop}%
\bibitem [{\citenamefont {{Gunn}}\ and\ \citenamefont
  {{Gott}}(1972)}]{Gunn1972}%
  \BibitemOpen
  \bibfield  {author} {\bibinfo {author} {\bibfnamefont {J.~E.}\ \bibnamefont
  {{Gunn}}}\ and\ \bibinfo {author} {\bibfnamefont {J.~R.}\ \bibnamefont
  {{Gott}}, \bibfnamefont {III}},\ }\href {\doibase 10.1086/151605} {\bibfield
  {journal} {\bibinfo  {journal} {\apj}\ }\textbf {\bibinfo {volume} {176}},\
  \bibinfo {pages} {1} (\bibinfo {year} {1972})}\BibitemShut {NoStop}%
\bibitem [{\citenamefont {{Press}}\ and\ \citenamefont
  {{Schechter}}(1974)}]{Press1974}%
  \BibitemOpen
  \bibfield  {author} {\bibinfo {author} {\bibfnamefont {W.~H.}\ \bibnamefont
  {{Press}}}\ and\ \bibinfo {author} {\bibfnamefont {P.}~\bibnamefont
  {{Schechter}}},\ }\href {\doibase 10.1086/152650} {\bibfield  {journal}
  {\bibinfo  {journal} {\apj}\ }\textbf {\bibinfo {volume} {187}},\ \bibinfo
  {pages} {425} (\bibinfo {year} {1974})}\BibitemShut {NoStop}%
\bibitem [{\citenamefont {{Huang}}\ \emph {et~al.}(2011)\citenamefont
  {{Huang}}, \citenamefont {{Bond}},\ and\ \citenamefont {{Kofman}}}]{HBK2011}%
  \BibitemOpen
  \bibfield  {author} {\bibinfo {author} {\bibfnamefont {Z.}~\bibnamefont
  {{Huang}}}, \bibinfo {author} {\bibfnamefont {J.~R.}\ \bibnamefont {{Bond}}},
  \ and\ \bibinfo {author} {\bibfnamefont {L.}~\bibnamefont {{Kofman}}},\
  }\href {\doibase 10.1088/0004-637X/726/2/64} {\bibfield  {journal} {\bibinfo
  {journal} {\apj}\ }\textbf {\bibinfo {volume} {726}},\ \bibinfo {eid} {64}
  (\bibinfo {year} {2011})},\ \Eprint {http://arxiv.org/abs/1007.5297}
  {arXiv:1007.5297 [astro-ph.CO]} \BibitemShut {NoStop}%
\bibitem [{\citenamefont {{Miao}}\ and\ \citenamefont
  {{Huang}}(2018)}]{MH2018}%
  \BibitemOpen
  \bibfield  {author} {\bibinfo {author} {\bibfnamefont {H.}~\bibnamefont
  {{Miao}}}\ and\ \bibinfo {author} {\bibfnamefont {Z.}~\bibnamefont
  {{Huang}}},\ }\href {\doibase 10.3847/1538-4357/aae523} {\bibfield  {journal}
  {\bibinfo  {journal} {\apj}\ }\textbf {\bibinfo {volume} {868}},\ \bibinfo
  {eid} {20} (\bibinfo {year} {2018})},\ \Eprint
  {http://arxiv.org/abs/1803.07320} {arXiv:1803.07320 [astro-ph.CO]}
  \BibitemShut {NoStop}%
\bibitem [{\citenamefont {{Huang}}(2021)}]{Huang2021}%
  \BibitemOpen
  \bibfield  {author} {\bibinfo {author} {\bibfnamefont {Z.}~\bibnamefont
  {{Huang}}},\ }\href {\doibase 10.1103/PhysRevD.104.103533} {\bibfield
  {journal} {\bibinfo  {journal} {\prd}\ }\textbf {\bibinfo {volume} {104}},\
  \bibinfo {eid} {103533} (\bibinfo {year} {2021})},\ \Eprint
  {http://arxiv.org/abs/2108.06089} {arXiv:2108.06089 [astro-ph.CO]}
  \BibitemShut {NoStop}%
\bibitem [{\citenamefont {{Abdul Karim}}\ \emph {et~al.}(2025)\citenamefont
  {{Abdul Karim}}, \citenamefont {{Aguilar}}, \citenamefont {{Ahlen}} \emph
  {et~al.}}]{DESIDR2}%
  \BibitemOpen
  \bibfield  {author} {\bibinfo {author} {\bibfnamefont {M.}~\bibnamefont
  {{Abdul Karim}}}, \bibinfo {author} {\bibfnamefont {J.}~\bibnamefont
  {{Aguilar}}}, \bibinfo {author} {\bibfnamefont {S.}~\bibnamefont {{Ahlen}}},
  \emph {et~al.},\ }\href {\doibase 10.1103/tr6y-kpc6} {\bibfield  {journal}
  {\bibinfo  {journal} {\prd}\ }\textbf {\bibinfo {volume} {112}},\ \bibinfo
  {eid} {083515} (\bibinfo {year} {2025})},\ \Eprint
  {http://arxiv.org/abs/2503.14738} {arXiv:2503.14738 [astro-ph.CO]}
  \BibitemShut {NoStop}%
\bibitem [{\citenamefont {{Aghanim}}\ \emph
  {et~al.}(2020{\natexlab{b}})\citenamefont {{Aghanim}}, \citenamefont
  {{Akrami}}, \citenamefont {{Ashdown}} \emph {et~al.}}]{Planck2018Like}%
  \BibitemOpen
  \bibfield  {author} {\bibinfo {author} {\bibfnamefont {N.}~\bibnamefont
  {{Aghanim}}}, \bibinfo {author} {\bibfnamefont {Y.}~\bibnamefont {{Akrami}}},
  \bibinfo {author} {\bibfnamefont {M.}~\bibnamefont {{Ashdown}}},  \emph
  {et~al.},\ }\href {\doibase 10.1051/0004-6361/201936386} {\bibfield
  {journal} {\bibinfo  {journal} {\aap}\ }\textbf {\bibinfo {volume} {641}},\
  \bibinfo {eid} {A5} (\bibinfo {year} {2020}{\natexlab{b}})},\ \Eprint
  {http://arxiv.org/abs/1907.12875} {arXiv:1907.12875 [astro-ph.CO]}
  \BibitemShut {NoStop}%
\bibitem [{\citenamefont {{Rosenberg}}\ \emph {et~al.}(2022)\citenamefont
  {{Rosenberg}}, \citenamefont {{Gratton}},\ and\ \citenamefont
  {{Efstathiou}}}]{Rosenberg2022}%
  \BibitemOpen
  \bibfield  {author} {\bibinfo {author} {\bibfnamefont {E.}~\bibnamefont
  {{Rosenberg}}}, \bibinfo {author} {\bibfnamefont {S.}~\bibnamefont
  {{Gratton}}}, \ and\ \bibinfo {author} {\bibfnamefont {G.}~\bibnamefont
  {{Efstathiou}}},\ }\href {\doibase 10.1093/mnras/stac2744} {\bibfield
  {journal} {\bibinfo  {journal} {\mnras}\ }\textbf {\bibinfo {volume} {517}},\
  \bibinfo {pages} {4620} (\bibinfo {year} {2022})},\ \Eprint
  {http://arxiv.org/abs/2205.10869} {arXiv:2205.10869 [astro-ph.CO]}
  \BibitemShut {NoStop}%
\bibitem [{\citenamefont {{Carron}}\ \emph {et~al.}(2022)\citenamefont
  {{Carron}}, \citenamefont {{Mirmelstein}},\ and\ \citenamefont
  {{Lewis}}}]{Carron2022}%
  \BibitemOpen
  \bibfield  {author} {\bibinfo {author} {\bibfnamefont {J.}~\bibnamefont
  {{Carron}}}, \bibinfo {author} {\bibfnamefont {M.}~\bibnamefont
  {{Mirmelstein}}}, \ and\ \bibinfo {author} {\bibfnamefont {A.}~\bibnamefont
  {{Lewis}}},\ }\href {\doibase 10.1088/1475-7516/2022/09/039} {\bibfield
  {journal} {\bibinfo  {journal} {\jcap}\ }\textbf {\bibinfo {volume} {2022}},\
  \bibinfo {eid} {039} (\bibinfo {year} {2022})},\ \Eprint
  {http://arxiv.org/abs/2206.07773} {arXiv:2206.07773 [astro-ph.CO]}
  \BibitemShut {NoStop}%
\bibitem [{\citenamefont {{Popovic}}\ \emph {et~al.}(2026)\citenamefont
  {{Popovic}}, \citenamefont {{Shah}}, \citenamefont {{Kenworthy}} \emph
  {et~al.}}]{DES-Dovekie}%
  \BibitemOpen
  \bibfield  {author} {\bibinfo {author} {\bibfnamefont {B.}~\bibnamefont
  {{Popovic}}}, \bibinfo {author} {\bibfnamefont {P.}~\bibnamefont {{Shah}}},
  \bibinfo {author} {\bibfnamefont {W.~D.}\ \bibnamefont {{Kenworthy}}},  \emph
  {et~al.},\ }\href {\doibase 10.1093/mnras/stag632} {\bibfield  {journal}
  {\bibinfo  {journal} {\mnras}\ }\textbf {\bibinfo {volume} {548}},\ \bibinfo
  {eid} {stag632} (\bibinfo {year} {2026})},\ \Eprint
  {http://arxiv.org/abs/2511.07517} {arXiv:2511.07517 [astro-ph.CO]}
  \BibitemShut {NoStop}%
\bibitem [{\citenamefont {Torrado}\ and\ \citenamefont
  {Lewis}(2021)}]{Torrado2020}%
  \BibitemOpen
  \bibfield  {author} {\bibinfo {author} {\bibfnamefont {J.}~\bibnamefont
  {Torrado}}\ and\ \bibinfo {author} {\bibfnamefont {A.}~\bibnamefont
  {Lewis}},\ }\href {\doibase 10.1088/1475-7516/2021/05/057} {\bibfield
  {journal} {\bibinfo  {journal} {JCAP}\ }\textbf {\bibinfo {volume} {05}},\
  \bibinfo {pages} {057} (\bibinfo {year} {2021})},\ \Eprint
  {http://arxiv.org/abs/2005.05290} {arXiv:2005.05290 [astro-ph.IM]}
  \BibitemShut {NoStop}%
\bibitem [{\citenamefont {Lewis}\ and\ \citenamefont
  {Bridle}(2002)}]{Lewis2002}%
  \BibitemOpen
  \bibfield  {author} {\bibinfo {author} {\bibfnamefont {A.}~\bibnamefont
  {Lewis}}\ and\ \bibinfo {author} {\bibfnamefont {S.}~\bibnamefont {Bridle}},\
  }\href {\doibase 10.1103/PhysRevD.66.103511} {\bibfield  {journal} {\bibinfo
  {journal} {Phys. Rev.}\ }\textbf {\bibinfo {volume} {D66}},\ \bibinfo {pages}
  {103511} (\bibinfo {year} {2002})},\ \Eprint
  {http://arxiv.org/abs/astro-ph/0205436} {arXiv:astro-ph/0205436 [astro-ph]}
  \BibitemShut {NoStop}%
\bibitem [{\citenamefont {{Neal}}(2005)}]{Neal2005}%
  \BibitemOpen
  \bibfield  {author} {\bibinfo {author} {\bibfnamefont {R.~M.}\ \bibnamefont
  {{Neal}}},\ }\href {https://arxiv.org/abs/math/0502099} {\bibfield  {journal}
  {\bibinfo  {journal} {ArXiv Mathematics e-prints}\ } (\bibinfo {year}
  {2005})},\ \Eprint {http://arxiv.org/abs/math/0502099} {math/0502099}
  \BibitemShut {NoStop}%
\bibitem [{\citenamefont {Lewis}(2013)}]{Lewis2013}%
  \BibitemOpen
  \bibfield  {author} {\bibinfo {author} {\bibfnamefont {A.}~\bibnamefont
  {Lewis}},\ }\href {\doibase 10.1103/PhysRevD.87.103529} {\bibfield  {journal}
  {\bibinfo  {journal} {Phys. Rev.}\ }\textbf {\bibinfo {volume} {D87}},\
  \bibinfo {pages} {103529} (\bibinfo {year} {2013})},\ \Eprint
  {http://arxiv.org/abs/1304.4473} {arXiv:1304.4473 [astro-ph.CO]} \BibitemShut
  {NoStop}%
\bibitem [{\citenamefont {Lewis}\ \emph {et~al.}(2000)\citenamefont {Lewis},
  \citenamefont {Challinor},\ and\ \citenamefont {Lasenby}}]{Lewis1999}%
  \BibitemOpen
  \bibfield  {author} {\bibinfo {author} {\bibfnamefont {A.}~\bibnamefont
  {Lewis}}, \bibinfo {author} {\bibfnamefont {A.}~\bibnamefont {Challinor}}, \
  and\ \bibinfo {author} {\bibfnamefont {A.}~\bibnamefont {Lasenby}},\ }\href
  {\doibase 10.1086/309179} {\bibfield  {journal} {\bibinfo  {journal}
  {Astrophys. J.}\ }\textbf {\bibinfo {volume} {538}},\ \bibinfo {pages} {473}
  (\bibinfo {year} {2000})},\ \Eprint {http://arxiv.org/abs/astro-ph/9911177}
  {arXiv:astro-ph/9911177 [astro-ph]} \BibitemShut {NoStop}%
\bibitem [{\citenamefont {Howlett}\ \emph {et~al.}(2012)\citenamefont
  {Howlett}, \citenamefont {Lewis}, \citenamefont {Hall},\ and\ \citenamefont
  {Challinor}}]{Howlett2012}%
  \BibitemOpen
  \bibfield  {author} {\bibinfo {author} {\bibfnamefont {C.}~\bibnamefont
  {Howlett}}, \bibinfo {author} {\bibfnamefont {A.}~\bibnamefont {Lewis}},
  \bibinfo {author} {\bibfnamefont {A.}~\bibnamefont {Hall}}, \ and\ \bibinfo
  {author} {\bibfnamefont {A.}~\bibnamefont {Challinor}},\ }\href {\doibase
  10.1088/1475-7516/2012/04/027} {\bibfield  {journal} {\bibinfo  {journal}
  {JCAP}\ }\textbf {\bibinfo {volume} {1204}},\ \bibinfo {pages} {027}
  (\bibinfo {year} {2012})},\ \Eprint {http://arxiv.org/abs/1201.3654}
  {arXiv:1201.3654 [astro-ph.CO]} \BibitemShut {NoStop}%
\bibitem [{\citenamefont {{Chevallier}}\ and\ \citenamefont
  {{Polarski}}(2001)}]{Chevallier01}%
  \BibitemOpen
  \bibfield  {author} {\bibinfo {author} {\bibfnamefont {M.}~\bibnamefont
  {{Chevallier}}}\ and\ \bibinfo {author} {\bibfnamefont {D.}~\bibnamefont
  {{Polarski}}},\ }\href {\doibase 10.1142/S0218271801000822} {\bibfield
  {journal} {\bibinfo  {journal} {International Journal of Modern Physics D}\
  }\textbf {\bibinfo {volume} {10}},\ \bibinfo {pages} {213} (\bibinfo {year}
  {2001})},\ \Eprint {http://arxiv.org/abs/gr-qc/0009008} {gr-qc/0009008}
  \BibitemShut {NoStop}%
\bibitem [{\citenamefont {{Linder}}(2003)}]{Linder03}%
  \BibitemOpen
  \bibfield  {author} {\bibinfo {author} {\bibfnamefont {E.~V.}\ \bibnamefont
  {{Linder}}},\ }\href {\doibase 10.1103/PhysRevLett.90.091301} {\bibfield
  {journal} {\bibinfo  {journal} {Physical Review Letters}\ }\textbf {\bibinfo
  {volume} {90}},\ \bibinfo {eid} {091301} (\bibinfo {year} {2003})},\ \Eprint
  {http://arxiv.org/abs/astro-ph/0208512} {astro-ph/0208512} \BibitemShut
  {NoStop}%
\bibitem [{\citenamefont {{Reischke}}\ \emph {et~al.}(2026)\citenamefont
  {{Reischke}}, \citenamefont {{St{\"o}lzner}}, \citenamefont {{Joachimi}}
  \emph {et~al.}}]{Reischke2026}%
  \BibitemOpen
  \bibfield  {author} {\bibinfo {author} {\bibfnamefont {R.}~\bibnamefont
  {{Reischke}}}, \bibinfo {author} {\bibfnamefont {B.}~\bibnamefont
  {{St{\"o}lzner}}}, \bibinfo {author} {\bibfnamefont {B.}~\bibnamefont
  {{Joachimi}}},  \emph {et~al.},\ }\href {\doibase
  10.1051/0004-6361/202558581} {\bibfield  {journal} {\bibinfo  {journal}
  {\aap}\ }\textbf {\bibinfo {volume} {709}},\ \bibinfo {eid} {A82} (\bibinfo
  {year} {2026})},\ \Eprint {http://arxiv.org/abs/2512.11041} {arXiv:2512.11041
  [astro-ph.CO]} \BibitemShut {NoStop}%
\bibitem [{\citenamefont {Akaike}(1974)}]{Akaike1974}%
  \BibitemOpen
  \bibfield  {author} {\bibinfo {author} {\bibfnamefont {H.}~\bibnamefont
  {Akaike}},\ }\href {\doibase 10.1109/TAC.1974.1100705} {\bibfield  {journal}
  {\bibinfo  {journal} {IEEE Transactions on Automatic Control}\ }\textbf
  {\bibinfo {volume} {AC-19}},\ \bibinfo {pages} {716} (\bibinfo {year}
  {1974})}\BibitemShut {NoStop}%
\end{thebibliography}

\end{document}